\pdfoutput=1
\documentclass[a4paper,11pt]{article}
\usepackage{pos}
\usepackage{booktabs}
\usepackage{multirow}

\renewcommand{\speakerText}{Speaker}

\graphicspath{{images/}}

\usepackage{float}
\usepackage[section]{placeins}
\usepackage{wrapfig}

\usepackage{orcidlink}
\newcommand{\orcid}[1]{\,\orcidlink{#1}}
\let\posPrintHeadAuthors\printHeadAuthors
\renewcommand{\printHeadAuthors}{%
  \begingroup\renewcommand{\orcid}[1]{}\posPrintHeadAuthors\endgroup}

\title{Cross-geometry transfer and model collapse in point cloud calorimeter shower generation}
\ShortTitle{Transfer and model collapse in point cloud shower generation}

\author[a,b]{Thorsten Buss\orcid{0000-0002-1717-2138}}
\author[b]{Frank Gaede\orcid{0000-0002-7055-9200}}
\author[a]{Gregor Kasieczka\orcid{0000-0003-3457-2755}}
\author*[a]{Lorenzo Valente\orcid{0009-0007-0080-8738}}
\author[a]{Duncan Weber}

\affiliation[a]{Universit\"at Hamburg, Institut f\"ur Experimentalphysik,\\
Luruper Chaussee 149, 22761 Hamburg, Germany}

\affiliation[b]{Deutsches Elektronen-Synchrotron DESY,\\
Notkestra\ss e 85, 22607 Hamburg, Germany}

\emailAdd{lorenzo.valente@uni-hamburg.de}

\abstract{%
Particle shower simulation is a major computational cost in high-energy physics. Monte Carlo methods such as \textsc{Geant4} are accurate but expensive, while most machine learning surrogates are tied to specific detector geometries and require retraining for each design change. We study cross-geometry transfer learning with CaloClouds~\textsc{II}, a generative model that produces point clouds rather than voxels and can project onto arbitrary detector readouts. We pre-train on photon showers in the International Large Detector (ILD) and adapt to electron showers in the cylindrical CaloChallenge Dataset~3. With only 100 target showers, fine-tuning reduces the geometric mean Wasserstein distance to \textsc{Geant4} by about 51\% over training from scratch. Bias-only fine-tuning (BitFit) stays within 5\% of full fine-tuning while updating only 17\% of the diffusion network parameters. We also examine model collapse in CaloClouds~\textsc{II} by retraining its normalising flow and diffusion model on its own generated showers across successive generations.%
}

\FullConference{23rd International Workshop on Advanced Computing and Analysis Techniques in Physics Research (ACAT2025)\\
8–12 September 2025\\
Hamburg, Germany\\}

\begin{document}
\maketitle

%

\section{Introduction}
Over the next decade, high-energy physics will produce far larger data volumes, driven by the High-Luminosity Large Hadron Collider (HL-LHC) and by high-granularity detectors with growing numbers of readout channels. \textsc{Geant4}~\cite{GEANT4:2002zbu} accurately reproduces the detector response, but a single HL-LHC event may take minutes of CPU time to simulate, and calorimeter shower development dominates that cost.

Machine learning fast simulation learns the final detector response directly from the incident particle information, instead of tracking each particle step by step through the material. It can reach speed-ups of several orders of magnitude.
Generative surrogates, such as generative adversarial networks, variational autoencoders, flows, diffusion and autoregressive models, now reproduce showers with high fidelity.
Most of these models, however, are tied to a single detector geometry and must be retrained from scratch for each new design or detector. Efficient adaptation across geometries remains an open problem.

Point cloud representations address this geometry dependence by describing a shower as a set of space points that project onto an arbitrary detector.
Flexibility alone, though, does not guarantee transfer, since moving between geometries also requires the shower physics learnt during pre-training.
In this work\footnote{This contribution summarises Ref.~\cite{Gaede:2025shc}, to which we refer for full details.} we study whether single-detector pre-training of the point cloud surrogate CaloClouds~\textsc{II}~\cite{Buhmann:2023kdg} can provide both the geometric flexibility and the transferable physics knowledge needed for cross-geometry adaptation.
In contrast to multi-detector grid pre-training, as in CaloDiT-2~\cite{Raikwar:2025fky}, we pre-train on photon showers in the planar International Large Detector (ILD) electromagnetic calorimeter (ECAL) and adapt to electron showers in the cylindrical CaloChallenge Dataset~3~\cite{Krause:2024avx}, comparing full fine-tuning with parameter-efficient fine-tuning (PEFT).

With as few as $10^2$ target showers, full fine-tuning reduces the geometric mean Wasserstein distance by about $51\%$ relative to training from scratch; the gains are largest in the low-data regime.
We present the first systematic comparison of parameter-efficient strategies for fast calorimeter simulation: bias-only fine-tuning (BitFit) remains within $5\%$ of full fine-tuning while updating only $17\%$ of the diffusion network's trainable parameters.
Because such surrogates may eventually retrain on their own synthetic output, we also examine model collapse: the behaviour of the normalising flow and diffusion model of CaloClouds~\textsc{II} under this iterative retraining.

\section{Datasets}
We study transfer between two shower datasets simulated with \textsc{Geant4}: ILD photons for pre-training and CaloChallenge Dataset~3 as the downstream target.

\paragraph{Pre-training dataset.}
The pre-training set comprises $524\,000$ photon showers~\cite{Buhmann:2023kdg} with energy uniform in $10$--$90$~GeV in the ECAL of the ILD, a linear collider detector concept. Its 30 layers alternate tungsten absorbers with $0.5$~mm silicon sensors, read out in $5\times5$~mm$^2$ cells.
The \textsc{Geant4} deposits are clustered by layer and projected onto a grid of 36 subcells per physical cell, giving compact point clouds of a few thousand points per shower.

\paragraph{Downstream dataset.}
Dataset~3 of the CaloChallenge~\cite{Krause:2024avx} contains electron showers with log-uniform energy from $1$~GeV to $1$~TeV in Par04, an idealised cylindrical calorimeter of concentric tungsten and silicon cylinders, scored along the shower axis in 45 layers of $18\times50=900$ voxels (radial $\times$ azimuthal) each, $40\,500$ voxels per shower.
These are converted to point clouds and aligned with the pre-training format through cylindrical smearing, sampling fraction reversal and point ordering~\cite{Gaede:2025shc}. We train on $100\,000$ showers and hold out $10\,000$ for validation and testing.

\section{Cross-Geometry Transfer Learning}
\begin{wrapfigure}{r}{0.5\linewidth}
  \centering
  \vspace{-\intextsep}
  \includegraphics[width=\linewidth]{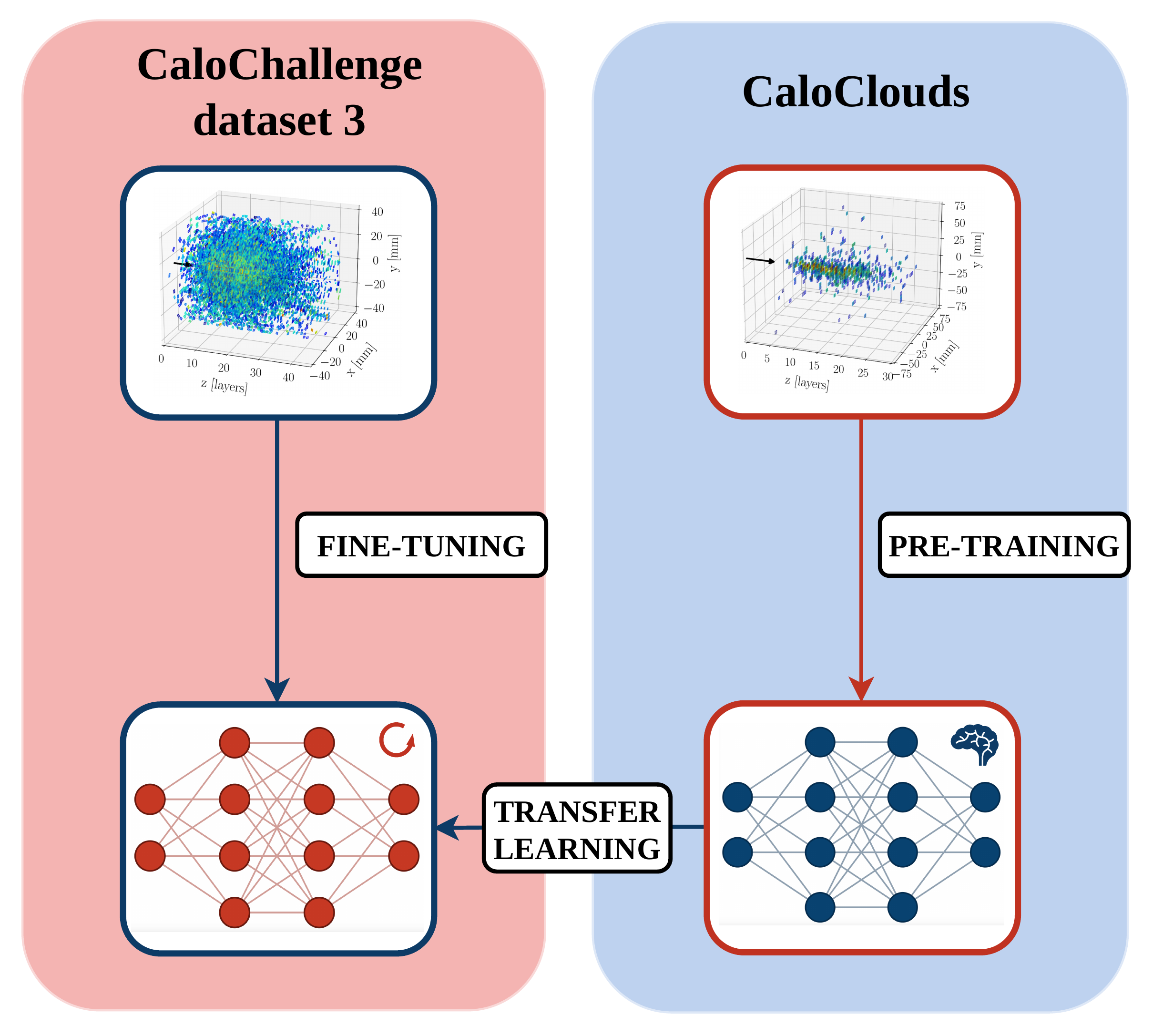}
  \caption{Overview of the cross-geometry transfer learning approach: a model pre-trained on ILD photon showers is fine-tuned to electron showers in the cylindrical CaloChallenge Dataset~3, in contrast to training from scratch.}
  \label{fig:overview}
\end{wrapfigure}
We build on CaloClouds~\textsc{II}~\cite{Buhmann:2023kdg}, a point cloud generative model that pairs a diffusion network (\textsc{PointWise~Net}), generating point positions and energies conditioned on the incident energy, with a normalising flow (\textsc{ShowerFlow}) for the per-layer point counts.
For the flow, we normalise the per-layer point counts on a fixed scale rather than per event, preserving cross-dataset scale information that per-event normalisation would otherwise compress.
We pre-train the model on ILD photons and adapt it to CaloChallenge electrons; the baseline is conventional training of each geometry from scratch.\footnote{The code for this study is available at \url{https://github.com/FLC-QU-hep/CaloTransfer}.} 
Because the model produces points rather than voxels, it can in principle adapt to any geometry (Fig.~\ref{fig:overview}), though we do not demonstrate this here; see Ref.~\cite{Gaede:2025shc}.

Adaptation is performed either by full fine-tuning or by parameter-efficient fine-tuning, which updates only a small subset of the diffusion network weights while the smaller flow is always fully fine-tuned. 
We study the bias-only BitFit~\cite{BenZaken:2021bitfit}, the Top2 scheme that updates the last two layers and the time-step embedding, and LoRA with rank 106~\cite{Hu:2021lora}. We evaluate all three across target dataset sizes $D \in \{10^2,\dots,10^5\}$ to probe data efficiency.
The transfer must bridge several changes at once: the geometry goes from planar rectangular cells to cylindrical radial and azimuthal voxels, the readout from 30 to 45 layers, and the incident particle from photons to electrons. The point clouds grow beyond three times the pre-training size at the highest energies, and the energy spectrum widens from a uniform $10$--$90$~GeV to a log-uniform $1$~GeV--$1$~TeV, well below and above the pre-training range.

\section{Results}
We present two sets of results: how efficiently the pre-trained surrogate transfers to a new detector geometry and how it behaves when repeatedly retrained on its own generated showers.

\begin{figure}[!ht]
  \centering
  \includegraphics[width=0.78\linewidth]{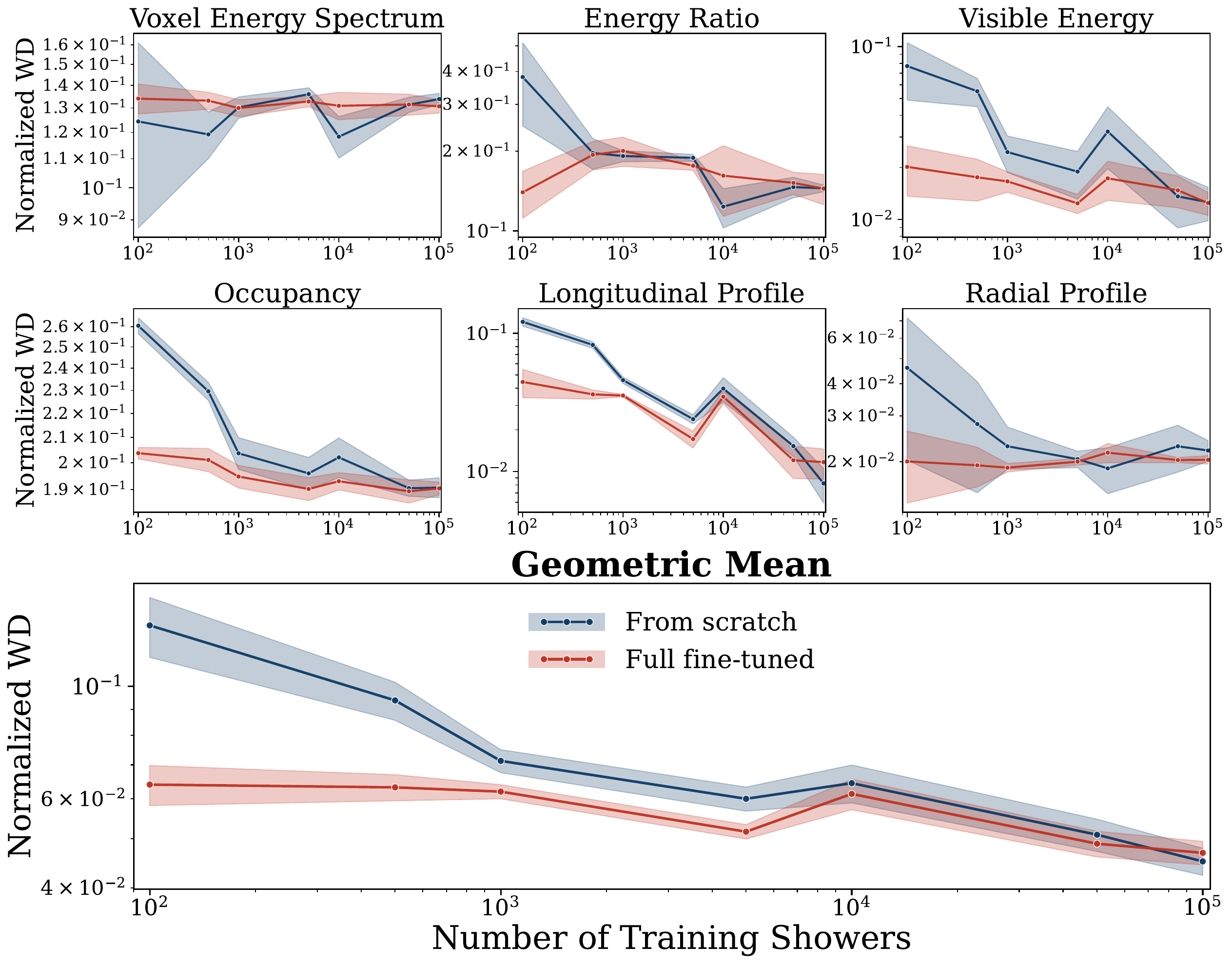}
  \caption{Wasserstein-1 distance between generated and \textsc{Geant4} showers as a function of the number of target domain training showers, for training from scratch and full fine-tuning, shown for each of the six shower observables and, in the final panel, their geometric mean.}
  \label{fig:transfer}
\end{figure}

\paragraph{Cross-geometry transfer.}
We quantify generation quality with six shower observables: the voxel energy spectrum, the ratio of measured to incident energy, the visible energy, the occupancy, and the longitudinal and radial profiles, each scored by the Wasserstein-1 distance to \textsc{Geant4} and combined by their geometric mean. Uncertainties are computed over five random seeds: RMS bands in Fig.~\ref{fig:transfer}, standard errors in Table~\ref{tab:peft}.
Transfer learning is most valuable when target data are limited: with only $10^2$ target showers, full fine-tuning reaches a geometric mean Wasserstein distance of $0.064\pm0.006$, against $0.132\pm0.018$ for training from scratch, a reduction of about $51\%$. The advantage of pre-training shrinks as more data become available (Fig.~\ref{fig:transfer}).
Among the PEFT strategies, the bias-only BitFit is the most effective, remaining within $5\%$ of full fine-tuning on the geometric mean (the Mean column of Table~\ref{tab:peft}) while updating only $17\%$ of the diffusion network parameters; Top2 is competitive using $44\%$; LoRA underperforms despite $52\%$.
We trace LoRA's difficulty to the high intrinsic dimensionality of the weight updates: some layers require ranks above $200$ (Ref.~\cite{Gaede:2025shc}). The success of bias-only BitFit instead points to a targeted recalibration of the existing weights.
A residual mismatch persists at $10^4$ samples in a few observables (Fig.~\ref{fig:transfer}), but is not visible in the aggregate geometric mean.

\begin{table}[!ht]
  \centering
  \caption{Geometric mean Wasserstein distance ($\times10^{-2}$, lower is better; best per column in bold) for each training strategy across target dataset sizes $D$ and arithmetically averaged over all sizes (Mean), with the fraction of updated diffusion network parameters. Uncertainties are standard errors over five seeds.}
  \label{tab:peft}
\small
\begin{tabular}{l|c|cccc|c}
\toprule
\multirow{2}{*}{\textbf{Method}} & \multirow{2}{*}{\textbf{Params (\%)}} & \multicolumn{4}{c|}{\textbf{Training Dataset Size}} & \multirow{2}{*}{\textbf{Mean}} \\
 & & 10$^2$ & 10$^3$ & 10$^4$ & 10$^5$ \\
\midrule
\textsc{From scratch} & 100 & 13.2±1.8 & 7.1±0.4 & 6.4±0.6 & \textbf{4.5±0.3} & 7.8±0.5 \\
\textbf{\textsc{Full fine-tuned}} & 100 & \textbf{6.4±0.6} & 6.2±0.2 & \textbf{6.1±0.4} & 4.7±0.3 & \textbf{5.9±0.2} \\ \midrule
\textsc{BitFit} & 17 & 6.9±0.4 & \textbf{6.1±0.2} & 6.2±0.5 & 5.4±0.3 & 6.2±0.2 \\
\textsc{Top2} & 44 & 7.6±0.6 & 6.2±0.2 & 6.5±0.4 & 7.1±0.8 & 6.8±0.3 \\
\textsc{LoRA R106} & 52 & 8.4±0.9 & 9.4±0.6 & 7.8±0.4 & 8.0±0.6 & 8.4±0.3 \\
\bottomrule
\end{tabular}
\end{table}

\begin{figure}[!ht]
  \centering
  \begin{minipage}{0.475\textwidth}\centering
    \begin{minipage}[b]{0.5\linewidth}\centering\includegraphics[height=1.8cm]{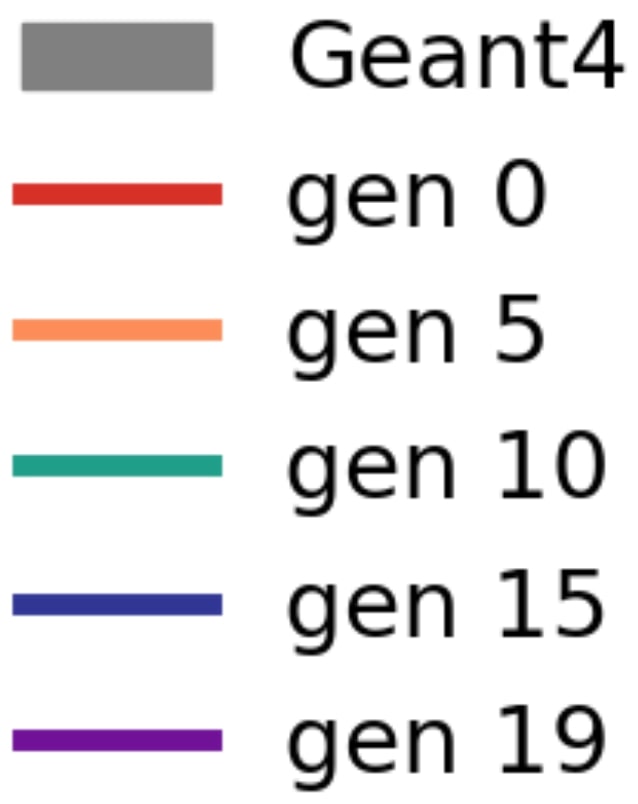}\end{minipage}%
    \begin{minipage}[b]{0.5\linewidth}\centering\includegraphics[height=1.35cm]{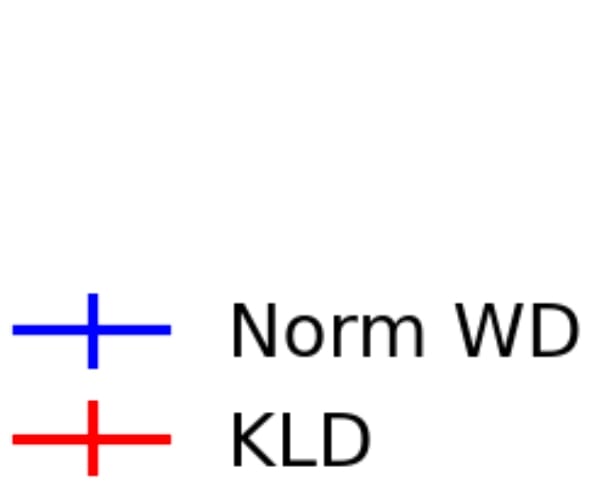}\end{minipage}
  \end{minipage}\hfill
  \begin{minipage}{0.475\textwidth}\centering
    \begin{minipage}[b]{0.5\linewidth}\centering\includegraphics[height=1.8cm]{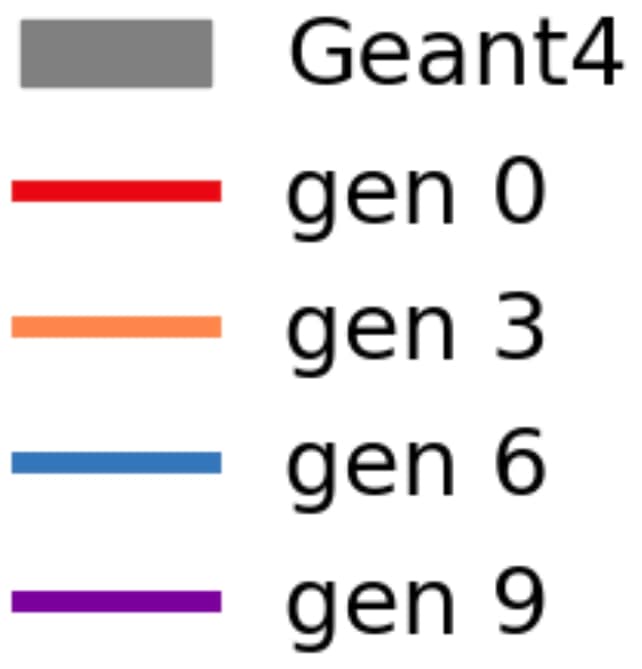}\end{minipage}%
    \begin{minipage}[b]{0.5\linewidth}\centering\includegraphics[height=1.35cm]{kld_legend2}\end{minipage}
  \end{minipage}

  \vspace{1mm}

  \begin{minipage}{0.475\textwidth}\centering
    \includegraphics[width=\linewidth]{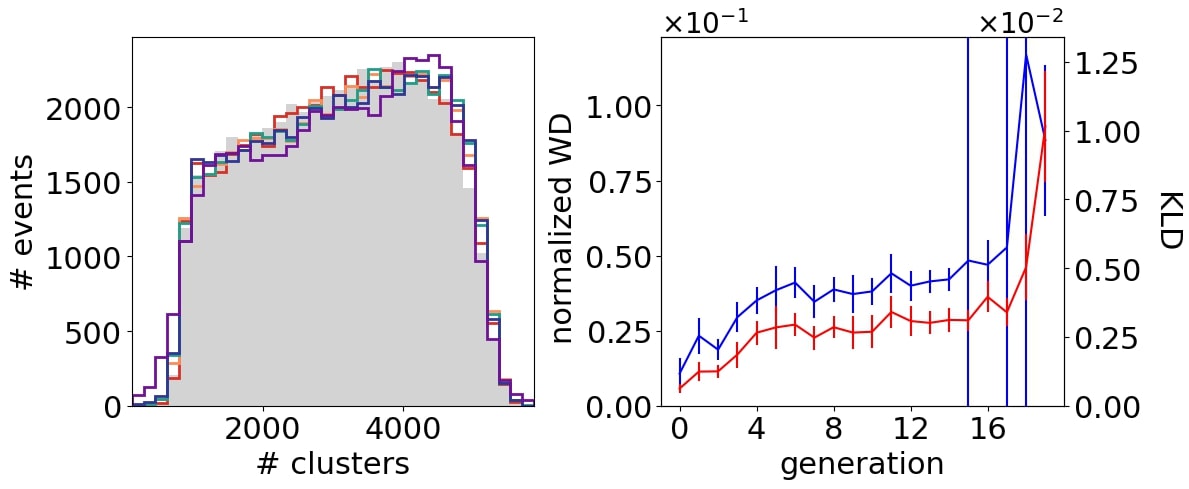}\\[2pt]
    \includegraphics[width=\linewidth]{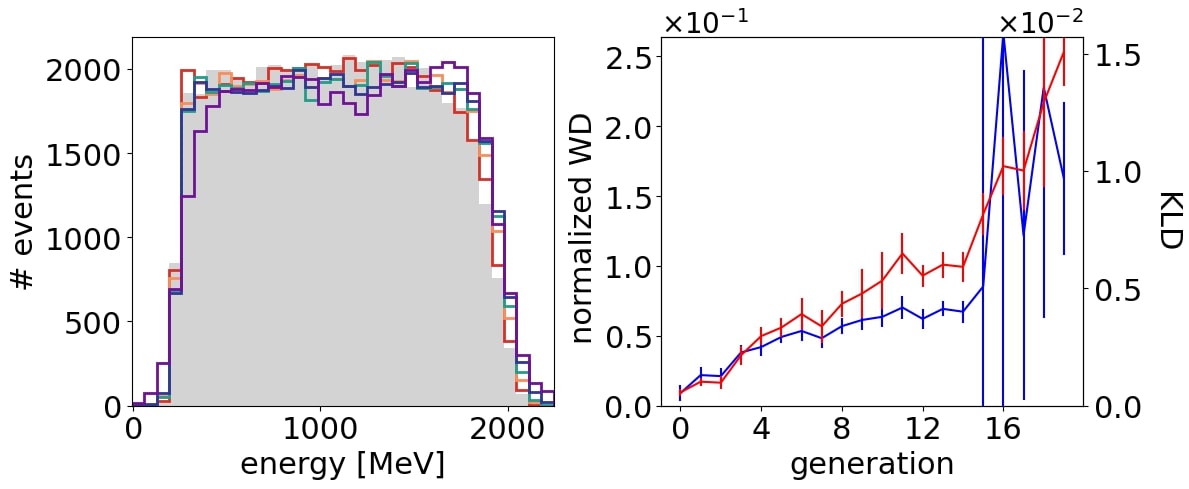}
  \end{minipage}\hfill
  \begin{minipage}{0.475\textwidth}\centering
    \includegraphics[width=\linewidth]{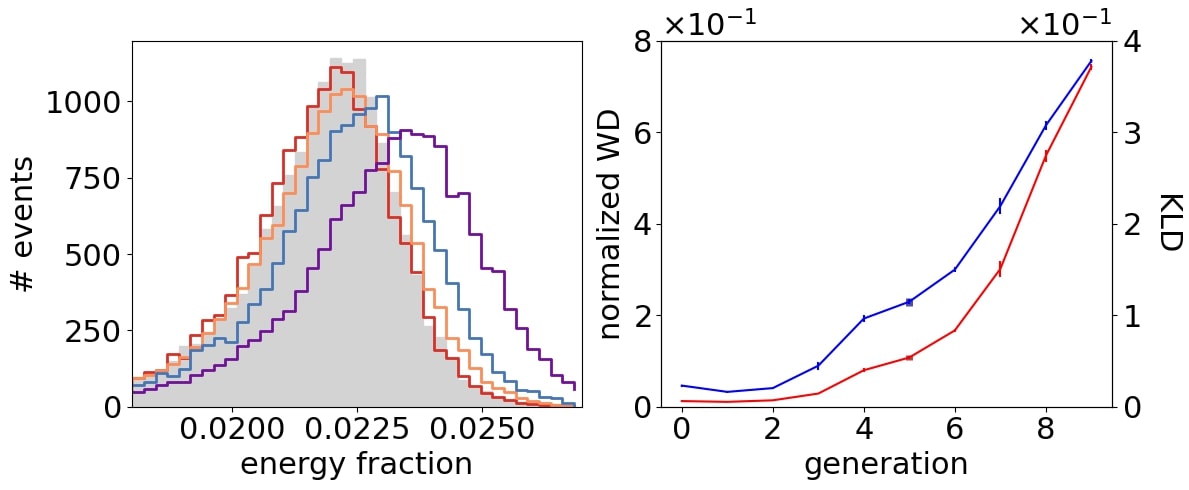}\\[2pt]
    \includegraphics[width=\linewidth]{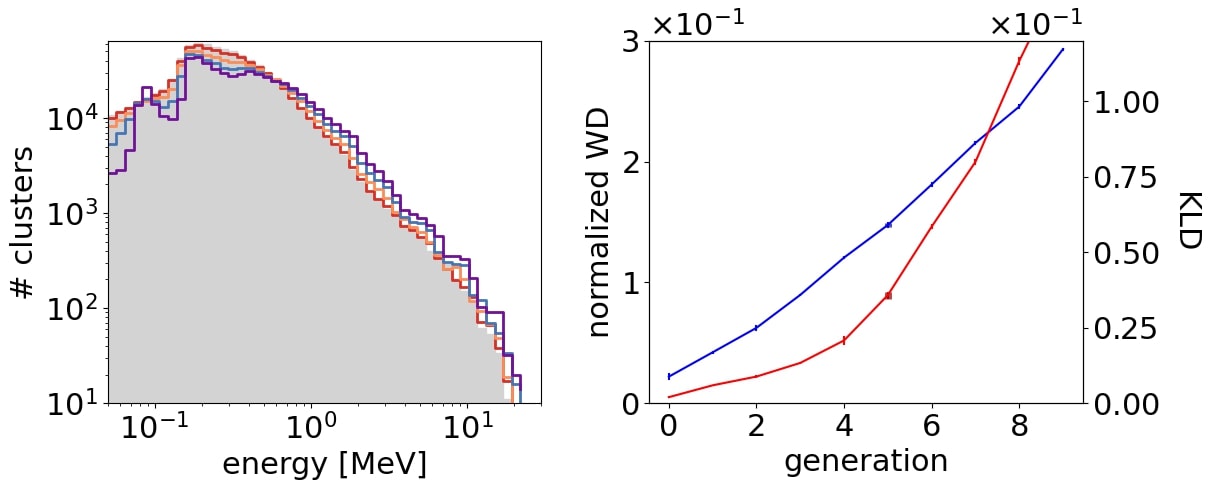}
  \end{minipage}
  \caption{Model collapse study under iterative retraining. \emph{Left} (\textsc{ShowerFlow}): number of clusters and total energy ($50\,000$ events); \emph{right} (diffusion): energy fraction and per-cluster energy ($16\,000$ events). For each observable, the left sub-plot is the histogram across generations (grey: \textsc{Geant4}) and the right the normalised Wasserstein distance and Kullback--Leibler divergence (KLD) to \textsc{Geant4} versus generation.}
  \label{fig:collapse}
\end{figure}

\paragraph{Iterative training.}
We examine the behaviour of the surrogate under iterative retraining in the ILD pre-training domain: each generation is trained on showers produced by the previous one in a self-consuming loop, separately for the diffusion and the normalising flow components of CaloClouds~\textsc{II}.
The results in Fig.~\ref{fig:collapse}, for four observables, show the onset of model collapse~\cite{Shumailov:2024}: the generated distributions degrade progressively relative to the reference, and for some observables the degradation accelerates in later generations.

\section{Conclusions and Outlook}
We have shown that single-detector pre-training on point clouds is a practical route to data-efficient cross-geometry transfer in calorimeter simulation: with only $10^2$ target showers it lowers the geometric mean Wasserstein distance to \textsc{Geant4} by about $51\%$ compared with training from scratch.
The advantage is largest in the low-data regime, where target showers are costliest to produce and full retraining is least affordable.
Most of this gain survives under parameter-efficient fine-tuning: BitFit matches full fine-tuning to within $5\%$ while training only $17\%$ of the diffusion network parameters.
We also quantify the onset of model collapse under iterative retraining, separately for the flow and diffusion components. The degradation grows over successive generations, indicating model collapse as a genuine failure mode for this point cloud surrogate.

The main limitation is that, without a direct comparison to multi-detector pre-training, we cannot benchmark the approach against foundation model strategies.
When multi-detector datasets are unavailable or computational resources are limited, single-detector pre-training is an adequate starting point for rapidly developing geometry-specific surrogates.
More broadly, these results show that meaningful transfer is achievable from a single geometry alone. Extending pre-training across diverse geometries and energies is a natural step towards a point cloud foundation model for calorimeter simulation.
Further directions include hadronic showers and mixed particle types.

\acknowledgments

This research was supported in part by the Maxwell computational resources operated at Deutsches
Elektronen-Synchrotron DESY, Hamburg, Germany, and by funding from the European Union\rq s
Horizon 2020 Research and Innovation programme (Grant Agreement No 101004761), the Deutsche
Forschungsgemeinschaft under Germany\rq s Excellence Strategy -- EXC 2121 Quantum Universe --
390833306, and the KISS consortium (05D23GU4, 13D22CH5) funded by the German Federal Ministry
of Research, Technology, and Space (BMFTR) in the ErUM-Data action plan.

\setlength{\bibsep}{3pt plus 1pt minus 1pt}
\bibliographystyle{JHEP}
\bibliography{biblio}

\end{document}